\documentclass[floatfix,altaffilletter,superscriptaddress,preprintnumbers, tightenlines,showpacs,showkeys,nofootinbib,notitlepage,twocolumn]{revtex4-2}\usepackage[utf8]{inputenc}
\usepackage{amssymb}
\usepackage{amsmath}
\usepackage{graphics}
\usepackage{graphicx}
\usepackage{tikz}
\usepackage[compat=1.0.0]{tikz-feynman}
\usepackage{xcolor}

\usepackage[caption=false]{subfig}

\newcommand{\beq}{\begin{equation}}
\newcommand{\eeq}{\end{equation}}

\newcommand{\dd}{\mathrm{d}}
\newcommand{\Tr}{\text{Tr}}

\newcommand{\proton}{\mathrm{p}}

\usepackage[normalem]{ulem}

\begin{document}
\title{Proton spin, topology and confinement: lessons from QCD$_2$}

\author{David Frenklakh}
\affiliation{Center for Nuclear Theory, Department of Physics and Astronomy, Stony Brook University, NY 11794-3800, USA}

\begin{abstract}
    We investigate the relation between the topology of a nucleon and its spin composition. We approach this question in (1+1) dimensional single-flavor QCD with a large number of color. In this limit the theory can be shown to be dual to the exactly solvable sine-Gordon model. The spectrum of baryons and mesons is known analytically, and the baryon is a topological kink of the sine-Gordon model. Using the method of solitonic constituents we construct the state of the baryon and extract its  $g_1$ structure function. Due to the topological nature of the baryon state this structure function is enhanced at low Bjorken $x$. We propose this enhancement as an experimental probe of the topological structure of the nucleon state.
\end{abstract}

\maketitle

\onecolumngrid

\begin{center}\small{Presented at DIS2022: XXIX International Workshop on Deep-Inelastic Scattering and Related Subjects, Santiago de Compostela, Spain, May 2-6 2022}
\end{center}

\vspace{0.6cm}

\twocolumngrid

\section{Introduction}

Since the establishment of QCD as the theory of strong interactions one problem remains unsolved: how to predict the physical properties of hadrons from the fundamental theory. One manifestation of this problem is the lack of understanding of the interplay between the nonperturbative structure of a baryon and its' spin decomposition among constituents, which is sometimes referred to as ``proton spin crisis" \cite{roberts_1990,Manohar:1992tz,Anselmino:1994gn,Filippone:2001ux,Bass:2004xa,Aidala:2012mv}. In particular, the role of topology of the baryon for its spin structure is not clearly identified yet.

We address this question in a simple toy model: QCD with the large number of colors (in the so-called 't Hooft limit) and one flavor in (1+1) spacetime dimensions. This model shares many important features with actual QCD in (3+1) dimensions: it has confinement, mass gap generation and its' spectrum includes baryons and mesons \cite{tHooft:1974pn,Witten:1983ar,Frishman:1992mr,Gross:1995bp}. In (1+1) dimensions there is no notion of spin, but there is still the notion of chirality, and axial current exists as well. This allows us to study chirality distributions, similar to those entering the proton spin decomposition in QCD$_4$.

At the same time, QCD$_2$ in the 't Hooft limit is dual to the exactly solvable Sine-Gordon model \cite{Coleman:1974bu,Faddeev:1977rm,Zamolodchikov:1978xm}. In this dual description, the baryon arises as a topological kink. We follow the approach of \cite{Dvali:2015jxa} to construct the quantum state of this kink as a coherent state of constituents carrying the information about the topology. This approach may be viewed as an improved version of parton model which accounts for the global topology of a baryon. We manage to evaluate chirality distribution in a baryon exactly and identify the contribution coming from its topological structure. 

Due to the format of the proceedings contribution the presentation has to be shortened and some interesting details omitted. We refer the interested reader to the original paper \cite{Florio:2022uvd} for more details and references.

\vspace{0.4cm}

\section{Motivation and background} \label{sec:prelim}

Proton spin crisis emerged when the results of EMC experiment on polarized deep inelastic scattering (pDIS) appeared. The analysis of pDIS starts with the antisymmetric part of the operator product expansion (OPE) of two electromagnetic current operators:

\onecolumngrid

\beq
W^{\mu\nu} = \frac{1}{4 \pi} \int \dd^4 x e^{iqx} \langle \mathrm{p},s|[j^\mu(x),j^\nu(0)]|\mathrm{p},s \rangle \underset{\substack{Q^2\to \infty }}{\sim} 2 \epsilon^{\mu\nu\lambda\rho}\frac{p_\lambda}{Q^2} \langle \mathrm{p},s|\left[C^{NS}\left(j_{\rho5}^3+\frac{1}{\sqrt{3}}j_{\rho5}^8\right)+\frac{2\sqrt{2}}{\sqrt{3}}C^{S}j^0_{\rho5}\right]|\mathrm{p},s \rangle
\eeq

\twocolumngrid

From the OPE one can derive a relation between the first moment of polarized structure function and axial charges of the nucleon:

\begin{align}
  \int_0^1 \dd x_Bg_1(x_B)  =  \frac{1}{12} a^{3} + \frac{1}{36} a^{8} + \frac{1}{9} a^{0} \ ,
\end{align}
where $g_1(x_B)$ is the polarized structure function and the axial charges are defined as

\begin{align}
a^3 = 2 g_A^3, \ \ \ \  a^8 = 2\sqrt{3} g_A^8, \ \ \ \ a^0 = \sqrt{6} g_A^0, \ \\
 2 M s^\mu g_A^{a} = \langle \proton,s| j_5^{\mu a}| \proton,s\rangle.~~~~~~~
\end{align}

Assuming $SU(3)$ flavor symmetry, the nucleon axial charges $a^3$ and $a^8$ are independently extracted from the data on neutron and hyperon decays \cite{Ellis:1973kp}. Therefore the measurement of the first moment of $g_1(x_B)$ leads to determining the singlet axial charge $a^0$. The matrix element of singlet axial current entering the definition of singlet axial charge cannot be calculated in QCD from first principles. At the same time, it is important to understand how the nonperturbative topological structure of the baryon state manifests itself in this matrix element. The goal of this work is to address this question in a simpler model, QCD in 1+1 dimensions.

The model used in this work contains a non-Abelian gauge field $A_{\mu}\in SU(N)$ and one flavor of fermion $q$ in the fundamental representation of $SU(N)$
\beq
\mathcal{L}_{QCD_2} = -\frac{1}{4} \Tr F_{\mu\nu}F^{\mu\nu} + i \bar{q} \gamma^\mu(\partial_\mu - i g A_\mu) q - m\bar{q} q \, , \label{eq:QCD2}
\eeq
where $F_{\mu\nu}$ the gluon field strength tensor and $m$ the mass of the fermion. We will work in the large $N$ 't Hooft limit, defined as
 
\begin{align}
    N\to\infty, \ \ g\to 0, \ \  N\cdot g^2\equiv \lambda = \text{const} \ .
\end{align}

In (1+1)-dimensional field theories there is a bosonization duality, relating theories with fermionic fields to bosonic theories and vice versa. QCD$_2$ also admits bosonization, and as was shown in  \cite{Steinhardt:1980ry} in the leading order in $\frac{1}{N}$ the corresponding bosonic model is the well-known sine-Gordon model of a single scalar field $\phi$:

\beq
\mathcal{L} = \frac{1}{2} (\partial_\mu \phi)^2 - m'^2\cos\left(\frac{\phi}{f}\right) \ . \label{eq:lagSG}
\eeq

Parameters of the sine-Gordon model, $f$ and $m'$ are inferred from the original theory parameters: 
 \beq
 f = \sqrt{\frac{N}{4\pi}}, ~~~~~~~ m' = \sqrt{\frac{N}{\sqrt{\pi}}mg}.
\eeq 

\begin{figure}
	\includegraphics[scale=0.185]{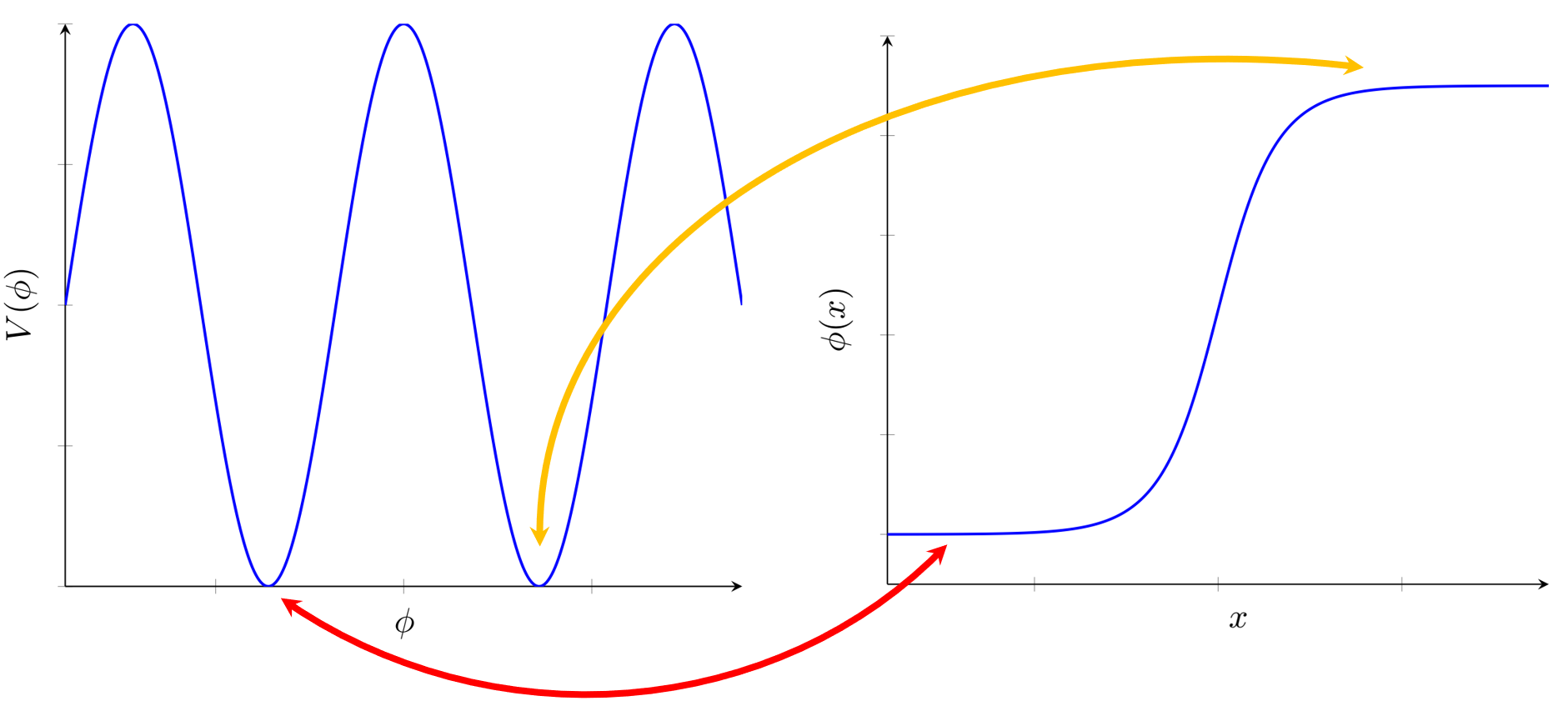}
	\caption{Left panel: potential of the sine-Gordon model. Right panel: spatial field profile of the kink solution. Arrows show identification of asymptotic values of the kink field with the two neighboring minima of the potential.}
	 \label{fig:kink}
\end{figure}

The most important property of the sine-Gordon model for us is that it admits certain classical solutions which are called topological kinks. Since the potential of the model is periodic in field $\phi$, there are infinitely many degenerate vacua. Therefore one can build a static finite-energy solution which at spatial $-\infty$ is at one vacuum and at spatial $+\infty$ is at the neighboring vacuum, as shown at Fig. \ref{fig:kink}. The explicit solution (for a stationary kink centered at $x=0$) is:

\beq \label{eq_kink}
\phi_c(x) = \sqrt{\frac{4 N}{\pi}} \arctan e^{\sqrt{\frac{4\pi}{N}}m' x} \ .
\eeq

Classical energy of the kink (since the kink is stationary, we will refer to it as the kink's mass) is

\beq \label{eq_m_kink}
M_{kink} = 4\sqrt{\frac{N}{\pi}} m' .
\eeq

To convince ourselves that the sine-Gordon kink is to be identified with the baryon, let us calculate its vector charge. In terms of the bosonic field $\phi$ the fermionic vector current $j^\mu = \bar{\psi}\gamma^\mu\psi$ is

\beq
j^\mu = \sqrt{\frac{N}{\pi}}\epsilon^{\mu\nu} \partial_\nu \phi \ ,
\eeq
therefore the total vector charge of the kink is

\beq
Q = \sqrt{\frac{N}{\pi}}\left[\phi(x\rightarrow\infty) - \phi(x\rightarrow -\infty)\right] = N\ .
\eeq

Since the charge is $N$, it means that the kink consists of $N$ quarks and hence represents the baryon in the bosonic language. 

Finally, since our goal is to calculate the matrix element of the axial current, it is useful to remember the connection between axial and vector currents in (1+1) dimensions and the corresponding bosonized version of the axial current:

\beq\label{vec_ax}
j^\mu_5 = \epsilon^{\mu\nu}j_\nu \Longrightarrow j^\mu_5 = \sqrt{\frac{N}{\pi}} \partial^\mu\phi.
\eeq

\section{Chirality distribution inside a kink in the Sine-Gordon model} \label{sec:results}

Our goal is to study the matrix element of axial current over a baryon state. As we have seen in the previous section, baryon in the bosonized version of QCD$_2$ is a topological kink, so we are going to compute

\beq
\langle kink|j_\mu^5|kink\rangle .
\eeq

First, we need to clarify what is meant by the quantum state $|kink\rangle$. Following the approach of \cite{Dvali:2015jxa} let us define the kink state such that 

\beq \label{eq_state}
\langle kink| \hat{\phi}(x) | kink \rangle = \phi_c (x),
\eeq
where $\hat\phi$ is the quantum field operator and $\phi_{c}$ is the kink profile \eqref{eq_kink}. Following the usual approach of canonical field quantization we can introduce a decomposition of the quantum field operator $\hat{\phi}$ into creation and annihilation operators along with a Fourier decomposition of the classical soliton profile $\phi_c(x)$:

\beq \label{eq:phi_decomp}
\hat{\phi}(x) = \int \frac{dk}{2\pi}\frac{1}{\sqrt{2\omega(k)}} ( a^{sol}_k e^{ikx} + a^{sol\dagger}_k e^{-ikx}), \
\eeq

\beq \label{eq_Fourier_coord}
\phi_c(x) = \int \frac{dk}{2\pi} \frac{1}{\sqrt{2\omega(k)}} (\alpha_k e^{ikx} + \alpha^*_k e^{-ikx}) \ .
\eeq

Here $a_k^{sol}$, $a_k^{sol\dagger}$ are correspondingly annihilation and creation operators of the so-called solitonic constituents, satisfying the usual commutation relations $[a_k^{sol}, a_{k'}^{sol\dagger}] = 2\pi\delta(k-k')$; $\alpha_k$ are Fourier coefficients of the classical soliton configuration.

Comparing the decompositions of the field operator and the classical field profile, we see that the definition of kink state (\ref{eq_state}) will be satisfied if
\beq
\label{eq:def_prop_a}
\langle kink| a^{sol}_k| kink \rangle = \alpha_k \ .
\eeq

Therefore we should define the kink state as a product of coherent states, each corresponding to a particular momentum:

\beq
|kink\rangle = \bigotimes_k |\alpha_k\rangle,
\eeq
where $|\alpha_k\rangle$ is a coherent state for momentum $k$, defined by $a_k^{sol} |\alpha_k\rangle = \alpha_k |\alpha_k\rangle$.

A useful way to further understand the structure of the coherent baryon state, first suggested in \cite{Dvali:2015jxa}, is to decompose the Fourier coefficients into a product of ``topology" and ``energy":

\beq
\alpha_ k  = t_k c_k ~~, ~~ t_k = \frac{i \sqrt{\omega(k)}}{k} ~~,~~ c_k = -\sqrt{\frac{\pi N}{2}}\frac{1}{\cosh{\sqrt{\frac{N}{4\pi}}\frac{\pi k}{2m'}}} \ . \nonumber
\eeq

The pole at $k=0$ means that there are infinitely many bosonic topological constituents at zero momentum, which are necessary to encode topology. The corresponding splitting in position space is a convolution:

\beq \label{eq_convolution}
\phi_c(x) = \sqrt{\frac{N}{\pi}} (\text{sign} \ast  \text{sech}) \left(\sqrt{\frac{N}{4\pi}}m' x\right)  + \frac{\pi}{2}  \ ,
\eeq
where the sign function encodes the topology of the field profile and  ``energy" is the rest.

The same procedure can be repeated for a kink  in the infinite momentum frame. Let us boost the kink with $\beta\rightarrow 1$, then the field profile becomes

\beq \label{eq:boosted_kink}
\phi_b (x,t)  \xrightarrow[\beta\rightarrow\ 1]{} \sqrt{\frac{N}{4\pi}} 4 \arctan e^{\sqrt{\frac{4\pi}{N}}m' \gamma \sqrt{2} x^+},
\eeq
with light-cone coordinates $x^{\pm} = \dfrac{1}{\sqrt{2}} (t\pm x)$, and $\gamma = \dfrac{1}{\sqrt{1-\beta^2}}$, as usual. 

Same as for a stationary kink, field operator can be decomposed into light-cone constituents and state of the kink constructed as a coherent state of these constituents. The constituent occupation number turns out to be

\beq
N_{k_+} = |\alpha_{k_+}|^2 = \frac{\pi N}{\sqrt{2} x_B p_+} \frac{1}{\cosh\left[ N x_B\right]^2},
\eeq
where we introduced $p_+ = \sqrt{2} \gamma M_{kink}$ and Bjorken $x_B = \dfrac{k_+}{p_+}$. The occupation number is divergent at small $x_B$, which once again is a manifestation of the kink's topology.

From the bosonization dictionary it follows that $j^\mu_5 = \sqrt{\dfrac{N}{\pi}} \partial^\mu \phi$ and therefore, in momentum space,

\beq \label{eq:ax_current}
 \langle kink |j_{5+}(x_B) | kink \rangle = N \frac{1}{\cosh\left[N x_B\right]} \ .
\eeq

\begin{figure}
    \includegraphics[scale=0.34]{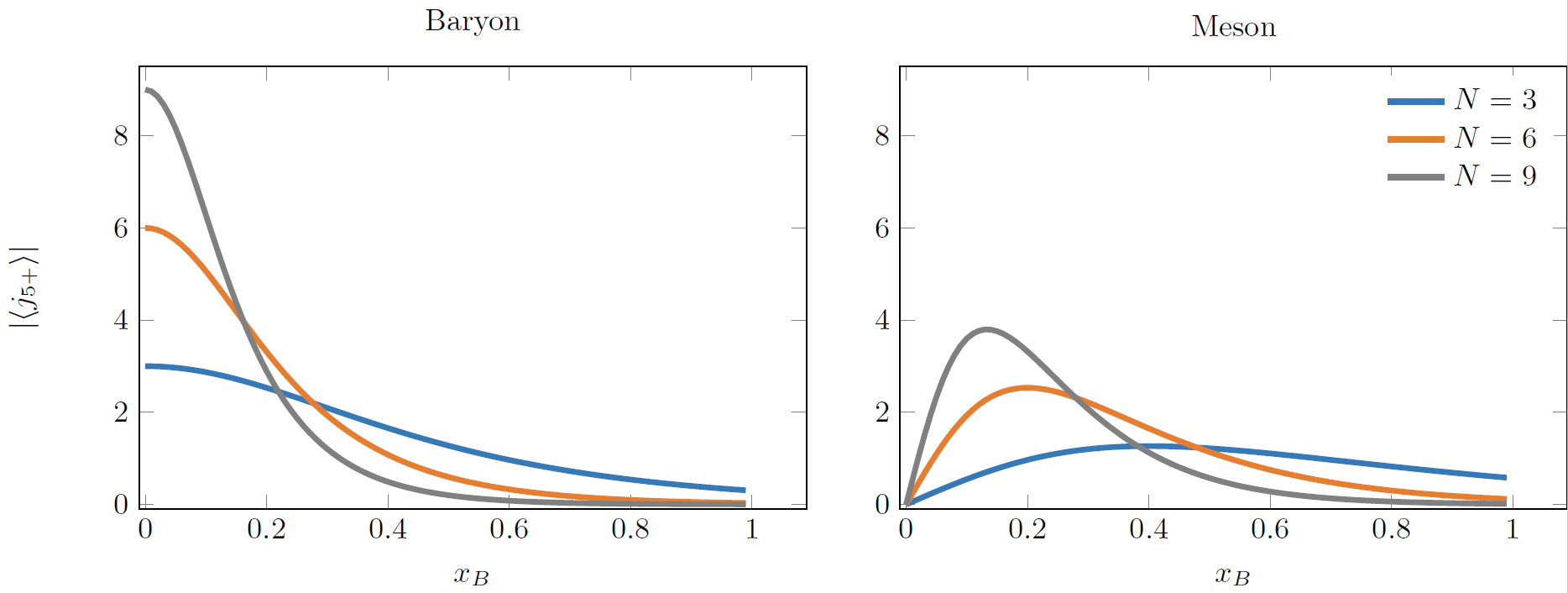}
    \caption{\textbf{Left panel:} The matrix element of the axial current inside a baryon as a function of Bjorken $x$ for different $N$. For large $N$, the $x_B\sim 1$ region is suppressed while the $x_B\rightarrow 0$ region is enhanced. 
    \textbf{Right panel:} Same as left panel, but for a ``meson".  The $x_B\rightarrow 0$ region contribution vanishes, reflecting the lack of baryon topology. }
    \label{plot:ax_current}
\end{figure}

As illustrated at Fig. \ref{plot:ax_current}, the matrix element of axial current is dominated by $x_B\sim\frac{1}{N}$. At large $N$ this means that the chirality distribution at small $x_B$ is enhanced while the region $x_B\sim 1$ is strongly suppressed.

In fact, our treatment allows to see that the enhancement of chirality distribution at small $x_B$ arises entirely from topology of the baryon. Topology can be factored out in the convolution form (\ref{eq_convolution}), which in the infinite momentum frame is

\beq
\phi_b (x^+) = \sqrt{\frac{N}{\pi}}(\text{sign} \ast \text{sech}) \left(\frac{\pi p_+ x^+}{2N}\right) \ .
\eeq

We introduce a nontopological state (``meson") by simply removing the sign function, so only the ``energy" part is left:

\beq
\phi_b^{n-t} (x^+) = \sqrt{\frac{N}{\pi}} \frac{1}{\cosh\left(\frac{\pi p_+ x^+}{2N}\right)} \ .
\eeq

This is not the actual meson of QCD$_2$ and is not a classical solution of the equations of motion, but a comparison to this state will help us clarify the role of topology for chirality distribution in a baryon. Chirality distribution in this ``meson" is calculated along the lines of the baryon calculation described above; as a result we get

\onecolumngrid

\twocolumngrid

\beq
\langle non-top.|j_{5+}(x_B)|non-top.\rangle (x_B) = \frac{2N^2 x_B i}{\pi \cosh(N x_B)} . \nonumber
\eeq

Chirality distributions in a baryon and in a ``meson" are compared on the left and right panels of  Fig. \ref{plot:ax_current}. We see that for a non-topological state chirality vanishes at $x_B\rightarrow 0$ which indicates that its enhancement in a baryon state is entirely a consequence of topology. At the same time, $x_B\sim 1$ region is not affected by the differences in topological structure.


\section{Discussion}
\label{sec:discussion}

In the conventional parton model the only place where the nonperturbative hadron structure enters parton distributions is through the initial conditions for the evolution equations. The perturbative QCD evolution is not sensitive to the nonperturbative hadron structure, and in particular it is the same for baryons and mesons. However, it is important to understand what are the effects of nonperturbative topological baryon structure on parton distributions. Here we addressed this question in the exactly solvable QCD$_2$ in the 't Hooft limit.

\vskip0.3cm

QCD$_2$ in the 't Hooft limit is dual to the sine-Gordon model and the baryons are described as topological kinks. Quantum state of such a kink can be constructed as a coherent state of solitonic constituents which carry the information about the kink's topology. We have calculated the matrix element of axial current over such a topological kink state and found that it is  enhanced at small Bjorken $x_B$. Moreover, the enhancement at very low $x_B$ originates entirely from the topology of the state, as for a non-topological ``meson" chirality distribution is drastically different: it vanishes at $x_B\rightarrow 0$.

\vskip0.3cm

All of our discussion took place in (1+1) dimensions, so it is natural to wonder whether the results are relevant to QCD$_4$. Indeed, the relation (\ref{vec_ax}) between axial and vector currents and most of the derivations of Section \ref{sec:results} rely heavily on the theory being (1+1)-dimensional. However, we believe that the results we obtained may be relevant in the real world. The first reason is that in QCD$_4$ there is the chiral anomaly \cite{Adler:1969gk,Bell:1969ts}, which connects the UV and IR \cite{Dolgov:1971ri}. Therefore, it makes polarized structure functions sensitive to the IR structure of the baryon, including its topological structure \cite{Jaffe:1989jz,Hatsuda:1988jv,Efremov:1989sn,Shore:1990zu,Tarasov:2020cwl,Tarasov:2021yll}. The second reason is the factorization between longitudinal and transverse degrees of freedom in high energy processes, which might make the $(1+1)$-dimensional dynamics considered above directly relevant in (3+1) dimensions.

\vskip0.3cm

In terms of phenomenology, our main prediction is the drastic difference  between spin distributions in baryons and mesons. We expect that the baryon topological structure leads to an enhancement of $g_1(x_B)$ structure function at small $x_B$ while for mesons this effect would be absent and $g_1$ would be suppressed at small $x_B$. Experimentally, this prediction can be tested though the diffractive DIS with a baryon in the target fragmentation region separated by a rapidity gap from the inelastic final state, which can access meson structure function. Lattice QCD provides another way to measure and compare polarized structure functions of baryons and mesons \cite{gockeler1996polarized,best1997pi,green2014nucleon,chen2016nucleon,lin2018parton}.

An interesting extension of this work would be to study chirality distributions in the case of QCD$_2$ with several flavors, and in particular study how they are related to the constituent quark picture discussed in \cite{Ellis:1992wu}. Another possible extension would be to study the $1/N$ corrections.

\section*{Acknowledgments} 
The author is grateful to D. Kharzeev and A. Florio for collaboration on this project which resulted in the paper \cite{Florio:2022uvd} as well as to V. Korepin for many discussions about the Sine-Gordon model. This work was supported by the U.S. Department of Energy,
Office of Science grant No.
DE-FG88ER41450


%

\end{document}